\documentclass[aps,prl,twocolumn,showpacs,nofootinbib]{revtex4}%
\usepackage{amsfonts}
\usepackage{amsmath}
\usepackage{amssymb}
\usepackage{epsfig}
\usepackage{graphicx}
\usepackage{dcolumn}%
\usepackage{bm}

\begin{document}

\title{Oddballs and a Low Odderon Intercept}
\author{Felipe J. Llanes-Estrada$^1$}
\altaffiliation[On leave at: ]{Theory Group, Stanford Linear 
Accelerator Center, 2571 Sand Hill Rd. 94025 Menlo Park CA}
\email{fllanes@fis.ucm.es}
\author{Pedro  Bicudo$^2$}
\email{bicudo@ist.utl.pt}
\author{Stephen R. Cotanch$^3$}
\email{cotanch@ncsu.edu}
\affiliation{$^1$Departamento de F\'{\i}sica Te\'orica I, Universidad
Complutense de Madrid, 28040 Madrid, Spain \\
$^2$Departamento de F\'{\i}sica and CFIF, 
Instituto Superior Tecnico, Av. Rovisco Pais, 1049-001  Lisboa, Portugal \\
$^3$Department of Physics, North Carolina State University,
Raleigh, NC 27695-8202}

\date{\today}

\begin{abstract}
We report an odderon Regge trajectory emerging from a field theoretical
Coulomb gauge QCD model for the odd signature $J^{PC}$ ($P=C=$ -1) 
glueball states (oddballs). The trajectory intercept  is clearly smaller
than the pomeron and even the $\omega$ trajectory's
intercept which provides an explanation for the nonobservation of the
odderon  in high energy scattering data.
To further support this result we compare to glueball lattice data and also
perform calculations with an alternative model based upon an exact Hamiltonian
diagonalization for three  constituent gluons. 
\end{abstract}


\maketitle


Regge trajectories \cite{Regge:1959mz} have 
long been an effective phenomenological tool in hadronic physics. 
In Regge theory the scattering amplitude is governed
by Regge poles, $\alpha_n(s)$, in the complex $J$ (angular momentum) plane. For 
integer $J$ the amplitude has a pole in the complex $s$ plane and, by crossing
symmetry, for $t < 0$ at high $s$   the cross section is dominated by the 
Regge trajectory, $\alpha(t) = bt + \alpha(0)$, with the
largest intercept,
$\alpha(0)$.  This conjecture provides a unifying connection between hadron
spectroscopy  (Chew-Frautschi plot of $J$ vs. $t = M_J^2$) and the 
high energy behavior of the total cross section which scales as $s^{1-\alpha(0)}$.
For elastic scattering the energy dependence is
well described by the leading Regge trajectory, the pomeron, having
$\alpha_P(0) \cong 1$ and
$b_P=0.2-0.3$ GeV$^{-2}$  (for recent fits see Ref. \cite{Pelaez:2004vs}).
Of course the pomeron does not relate to conventional hadron
spectra since  meson trajectories typically have  larger slopes, 
$b_M \cong .9$ GeV$^{-2}$, and smaller intercepts, $\alpha_M(0) \cong .55$. 
According to the  glueball-pomeron conjecture 
\cite{Simonov:1990uq,Llanes-Estrada:2000jw},
which is supported by lattice 
data \cite{Meyer:2004jc} and other models \cite{Brau:2004xw}, 
this trajectory is instead connected to
glueball spectroscopy.  Related,  
the different pomeron and meson trajectory slopes can be
generated~\cite{Llanes-Estrada:2000jw} by the gluon and quark color factors,
respectively, used in  confining potential
models. Due to the large gluon mass gap, which suppresses relativistic corrections
and transverse gluon exchange 
\cite{Alkofer:2004it}, these models tend to be more robust for glueballs than
mesons.  They produce a pomeron consisting of
even signature $J^{++}$ glueballs 
having maximum  intrinsic spin $S$ coupled to minimum possible
orbital
$L$.

Of active interest is   the odd signature, $P
= C =$ -1 counterpart to the pomeron,  the odderon~\cite{ln}, for
which there is no firm experimental evidence. 
Whereas the pomeron predicts asymptotically equal $pp$ and 
$\bar{p}p$ cross sections, the competitive presence of the odderon or any other $C
=$ -1 trajectory would produce a difference. However, high energy measurements
reveal a minimal difference indicating that the odderon, if it exists,
would have a smaller intercept probably at most comparable to the $\omega$ value,
$\alpha_{\omega}(0)=0.5$.
Indeed, dedicated exclusive searches at HERA~\cite{Adloff:2002dw} exclude  an
odderon Regge trajectory with an intercept greater than 0.7.  
Although perturbative QCD calculations~\cite{Ewerz:2003xi} based on the BKP 
equation predict an odderon intercept close to 1, 
they are only reliable for both 
$s, -t>>\Lambda_{QCD}$ and thus suspect for  
$\alpha(t=0)$. For example, the predicted pomeron intercept using the 
similar BFKL equation is 1.5 in conflict with data.

In this paper we summarize recent nonperturbative QCD calculations 
based on lattice gauge theory and QCD models  
incorporating   fundamental elements which produce  realistic hadron spectra
with Regge trajectories.
Our key results clarify and dispel several misconceptions concerning the
odderon.  First, we document an odderon trajectory of odd
signature $J^{--}$ glueballs (oddballs) with intercept below 0.7,
consistent with experimental searches.  Second, the odderon and pomeron have
similar slopes, in  analogy with the similar-slope  meson
(two-body) and baryon (three-body) Regge trajectories.  
Third, the lightest resonance on the odderon is the $3^{--}$,  not
the $1^{--}$ which is on a daughter trajectory.
In the framework of constituent models, the
odderon begins with a $3^{--}$ three gluon $L = 0$ state  
with maximum spin $S$, similar to the two gluon pomeron
beginning with the s wave, spin aligned $2^{++}$ glueball. 
Higher  odd signature  states on the odderon have the same $S$
but increasing $L$ (e.g. d, g ... waves producing $J^{PC}
= 5^{--},7^{--}$...).


The first constituent three gluon calculations 
\cite{cs1983,Hou:1982dy} utilized a non-relativistic
potential model. They  reported~\cite{Hou:1982dy} the nearly degenerate 
lightest states,
$J^{PC} = 0^{-+}$,
$1^{--}$  and $3^{--}$,  with masses about 4.8 times the
constituent gluon mass. Hyperfine splittings breaking this
degeneracy were considered in
the bag model
\cite{Barnes:1981kq} but the global  mass scale  was too low 
~\cite{Simonov:1990uq}.
Another problem confronting early constituent gluon models was a spurious two gluon
$J=1$ state (unlike photons constituent gluons have 3 spin projections) which
violates boson statistics (Yang's theorem).  This issue is naturally resolved in
the Coulomb gauge effective QCD
formulation~\cite{Szczepaniak:1995cw} in which the Fock operator commutation
relations permit only transverse gluons respecting  Yang's
theorem.  

Our calculation is based on this  relativistic many-body 
Hamiltonian formulation which unifies the quark and glue sectors. 
This approach dynamically generates constituent  gluon
and quark
masses (while respecting chiral symmetry~\cite{flscprl,flscnp}), produces
reasonable quark and gluon condensates, describes flavored meson spectra,
including hyperfine splittings~\cite{flscss}, and predicts exotic
hybrids~\cite{flscplb}  and $C = $+1
glueballs~\cite{Llanes-Estrada:2000jw,Szczepaniak:1995cw} consistent with lattice
gauge results. It is also noteworthy that this formulation entails only two
dynamical parameters (same for both quark and glue sectors).  Consult
Refs.~\cite{flscnp,flscss,sc} for further details. 

The effective QCD Hamiltonian
in the gluon sector is
\begin{eqnarray}
H^{g}_{eff} &=&  
Tr  \int d {\bf x}\left[ {\bm \Pi}^a({\bf x})\cdot {\bm \Pi}^a({\bf x}) 
+ {\bf B}_A^a({\bf x})\cdot{\bf B}_A^a({\bf x}) \right ]  \nonumber \\
&-& \frac {1} {2}  \int d {\bf x} d {\bf y}\thinspace
\rho^a_g({\bf x}) V({\bf x},{\bf y})  \rho^a_g({\bf y}) \ ,
\end{eqnarray}
with color charge density 
$\rho^a_g({\bf x}) = f^{abc}{\bf A}^b({\bf x})\cdot{\bf {\bm \Pi}}^c({\bf x})$,
gauge fields   ${\bf A}^a$, conjugate momenta ${\bm \Pi}^a = - {\bf
E}^a$ and Abelian components
${\bf B}^a_A = {\bm \nabla} \times {\bf A}^a$,   for $a = 1,2,...8$.
The  normal mode expansions are
\begin{eqnarray}
\label{colorfields2}
{\bf A}^a({\bf{x}}) &=&  \int
\frac{d{\bf{q}}}{(2\pi)^3}
\frac{1}{\sqrt{2\omega_k}}[{\bf a}^a({\bf{q}}) + {\bf a}^{a\dag}(-{\bf{q}})]
e^{i{\bf{q}}\cdot
{\bf {x}}} \ , \ \ \
\\ 
{\bm \Pi^a}({\bf{x}}) &=& -i \int \frac{d{\bf{q}}}{(2\pi)^3}
\sqrt{\frac{\omega_k}{2}}
[{\bf a}^a({\bf{q}})-{\bf a}^{a\dag}(-{\bf{q}})]e^{i{\bf{q}}\cdot
{\bf{x}}}  ,
\end{eqnarray}
with the Coulomb gauge transverse condition,  
${\bf q}\cdot {\bf a}^a ({\bf q}) = \\ (-1)^\mu q_{\mu} a_{-\mu} ^a ({\bf q}) =0$.
Here $a_{\mu}^a({\bf{q}})$ ($\mu = 0, \pm 1$)
are the bare  gluon Fock operators  from which, by a
Bogoliubov-Valatin canonical transformation, the dressed gluon or quasiparticle
operators, 
$\alpha^a_{\mu} ({\bf{q}}) = \cosh \Theta(q) \,
a_{\mu}^a({\bf{q}}) +
\sinh \Theta(q) \,  a_{\mu}^{a\dagger}(-{\bf{q}})$, emerge.
This  similarity transformation is a
hyperbolic rotation similar to the BCS fermion treatment. These operators excite
constituent gluon quaisparticles from the BCS vacuum, $|\Omega>_{\rm BCS}$,
and satisfy the  transverse commutation relations,
$[\alpha^a_{\mu}({\bf q}),\alpha^{b \dagger}_{\nu}({\bf q}')]=\delta_{ab}
(2\pi)^3 \delta^3({\bf q}-{\bf q}')D_{{\mu} {\nu}}({\bf q})  $,
with  $D_{{\mu} {\nu}}({\bf q}) = \left( 
\delta_{{\mu}{\nu}}- (-1)^{\mu}\frac{q_{\mu} q_{-\nu}}{q^2} \right)$.
Finally, the quasiparticle or gluon self-energy, 
$\omega(q) = q
e^{-2\Theta(q)}$, satisfies a gap equation 
\cite{Szczepaniak:1995cw}.
Both confinement and the leading QCD canonical interaction are
contained in the Cornell type  potential
$V =   \sigma r -\frac {\alpha_s} {r}$ with string tension,
$\sigma = 0.18$
GeV$^2$, determined by  lattice gauge calculations and $\alpha_s = 0.42$. For
this interaction the calculated gluon  constituent mass from the gap equation,
which uses the cut-off/renormalization parameter $\Lambda =$ 1.1 GeV, is
$m_g \equiv
\omega(0) \cong$ 0.8 GeV.  Previous two constituent applications ($q \bar{q}$
mesons and $gg$ glueballs)  involved Tamm-Dancoff (TDA) and Random Phase
 approximation (RPA) Hamiltonian diagonalizations whereas three-body ($q \bar{q}
g$  hybrids) predictions utilized a variational calculation which was found to be
accurate when tested in  two-body systems.  We therefore adopt the  variational
method for predicting the $ggg$ oddballs which is the minimum Fock space
assignment for $C =$ -1 glueballs. 
The three-gluon variational wavefunction  is
\begin{eqnarray}
\lefteqn{\hspace{.5cm}\arrowvert \Psi^{JPC} \rangle = \int d{\bf q}_1 d{\bf q}_2
d{\bf q}_3 
\delta({\bf q}_1 + {\bf q}_2 + {\bf q}_3 )} \\ 
& & 
\hspace{-.3cm}  
F^{JPC}_{\mu_1 \mu_2 \mu_3}({\bf q}_1,{\bf q}_2,{\bf q}_3) 
C^{abc}
\alpha^{a\dagger}_{\mu_1}({\bf q}_1)
\alpha^{b\dagger}_{\mu_2}({\bf q}_2) \alpha^{c\dagger}_{\mu_3}({\bf q}_3)
\arrowvert \Omega \rangle_{\rm BCS} \ , \nonumber
\end{eqnarray}
with  summation over repeated indices. The
color tensor $C^{abc}$  is either totally 
antisymmetric $f^{abc}$ (for $C =$ 1) or symmetric 
$d^{abc}$ (for $C =$ -1).
Boson statistics thus requires the $C =$ -1 oddballs to have a
symmetric space-spin wavefunction taken here to have form 
\begin{eqnarray} 
F^{JPC}_{\mu_1 \mu_2 \mu_3}({\bf q}_1,{\bf q}_2,{\bf q}_3) = [c_{12} f(q_1,q_2) +
c_{23} f(q_2,q_3) +
\nonumber \\
 c_{13} f(q_1,q_3)]
 [Y_L^{\lambda}(\hat{\bf q}_1) + Y_L^\lambda (\hat{\bf q}_2) + 
Y_L^{\lambda}(\hat{\bf q}_3)] \ ,
\label{eqf12} \\
c_{12} = \langle 1 \mu_1 1 \mu_2 \arrowvert s \mu_s \rangle
\langle s \mu_s 1 \mu_3 \arrowvert S \mu \rangle 
\langle L \lambda S\mu \arrowvert J
{\cal M}\rangle \ ,
\label{c12}
\end{eqnarray}
which is appropriate for this analysis of the lightest states
which are either s or d wave
oddballs with a single  $L$ excitation.
The other two coefficients in Eq. (\ref{eqf12}) can be obtained
by permuting the indices in Eq. (\ref{c12}).
A more comprehensive treatment would include
other orbital angular excitations but since these are considerably separated in
energy (more so than for  conventional meson states) this mixing is small. Similar
comments apply  to higher radial 
wavefunction components.
Several forms for the variational radial wavefunction, $f(q,q')$,
involving two variational parameters, $\beta$ and $\beta'$, were 
investigated including a separable form, $f(q) f(q')$, which only involves 
one parameter. 
From previous experience~\cite{flscplb}, reliable, accurate variational
solutions can be obtained if these  functions have a
bell-shaped form with scalable variational parameters.
In particular we found
numerical solutions of the  $\rho$ meson TDA problem for 
the same potential to be especially accurate.





\begin{figure}
\psfig{figure=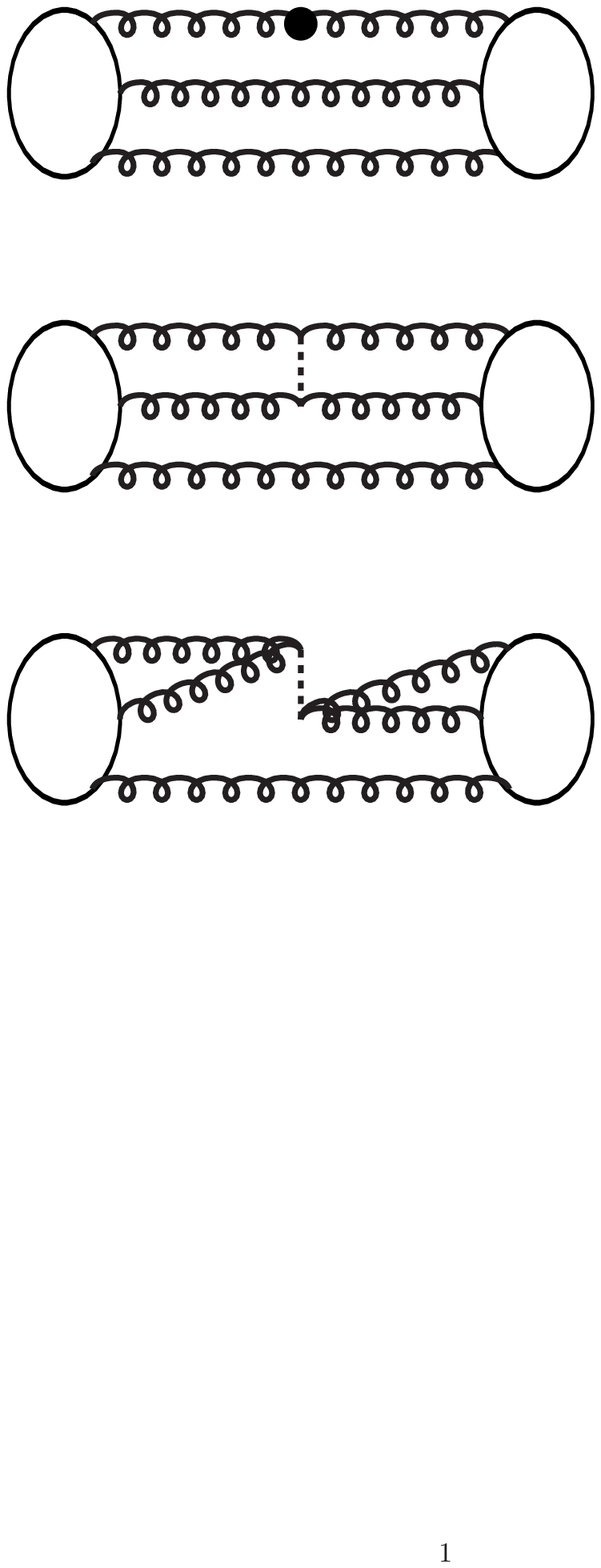,height=2.85in}
\caption{\label{TDA3g} Interaction diagrams for the $ggg$ 
state  in the $H^g_{eff}$ approach.
Top to bottom
corresponds to the gluonic self-energy, scattering and annihilation interactions,
respectively.}
\end{figure}

The variational equation for the $J^{PC}$ glueball with mass $M_{JPC}$ is  
\begin{equation}
\frac{\langle \Psi^{JPC} \arrowvert H^{g}_{eff} \arrowvert \Psi^{JPC} \rangle}
{\langle \Psi^{JPC}\arrowvert \Psi^{JPC} \rangle} = M_{JPC} \ .
\end{equation}
For clarity and insight it is worthwhile to express $H^{g}_{eff} =
H_{se}+H_{sc}+H_{an}$ as a sum of three terms corresponding
to self-energy
(two-body contributions), $H_{se}$,   scattering (instantaneous exchange
interaction between two gluons), $H_{sc}$, and  annihilation, $H_{an}$, of two
gluons. 
Figure~\ref{TDA3g} depicts the three-gluon diagram for each interaction.
The contribution from the self-energy term  is
\begin{eqnarray}
\lefteqn{\hspace{-.30cm}\frac{\langle \Psi \arrowvert H_{se} \arrowvert \Psi
\rangle} {\langle \Psi\arrowvert \Psi \rangle} =  18 \! \! \int \! \! d{\bf q}_1
d{\bf q}_2  
F^*_{\mu_1 \mu_2 \mu_3}({\bf q}_1,{\bf q}_2) F_{\nu_1 \nu_2 \nu_3}({\bf 
q}_1,{\bf q}_2) } \nonumber \\
& &\hspace{-.5cm}D_{\mu_1\alpha}({\bf q}_1) D_{\mu_2 \nu_2}({\bf q}_2)
D_{\mu_3 \nu_3}({\bf q}_3)D_{\beta \nu_1}({\bf q}_1) 
\left[ \frac{\omega_1^2+q_1^2}{2\omega_1}\delta_{\alpha\beta}  \right.
\nonumber  \\
& &\left. \hspace{1.5cm}
 - \, \,\frac{3}{4} \int \frac{d{\bf q}}{(2\pi)^3}
\hat{V}(q) \frac{\omega_1^2+\omega_6^2}{\omega_1\omega_6}
D_{\alpha \beta}({\bf q}_6) 
\right] ,  
\end{eqnarray} 
where we have suppressed the $J^{PC}$ superscript and the dependent
variable ${\bf q}_3 = -{\bf q}_1 - {\bf q}_2$. Again
$\omega_i=\omega({\bf q}_i)$ is the solution to the gluon 
gap equation \cite{Llanes-Estrada:2000jw,flscnp},  
${\bf q}_6 ={\bf q}_1 + {\bf q} $ and ${\bf q}$ is the  momentum transferred
by the  interaction. 
Similarly the scattering contribution is 
\begin{eqnarray}
\lefteqn{\hspace{-.10cm}
\frac{\langle \Psi \arrowvert H_{sc} \arrowvert \Psi \rangle}
{\langle \Psi\arrowvert \Psi \rangle} =  -\frac{9}{2}
{\cal{F}}_{sc}^{\pm}\int \! d{\bf q}_1 d{\bf q}_2 d{\bf q}
F^*_{\mu_1 \mu_2 \mu_3}({\bf q}_1,{\bf q}_2)\hat{V}(q)} \nonumber
\\
& &\hspace{-.4cm}F_{\nu_1 \nu_2 \nu_3}({\bf q}_4,{\bf q}_5) 
 \frac{(\omega_1+\omega_4)(\omega_2+\omega_5)}
{\sqrt{\omega_1\omega_2\omega_4\omega_5}}
D_{\mu_3 \nu_3}({\bf q}_3) D_{\mu_2 \beta}({\bf q}_2) 
\nonumber
\\
& & \hspace{2.5cm}
D_{\beta \nu_2}({\bf q}_5) D_{\mu_1 \alpha}({\bf q}_1)
D_{\alpha \nu_1}({\bf q}_4) \ , 
\end{eqnarray}
with ${\bf q}_4 ={\bf q}_1 - {\bf q} $, ${\bf q}_5 ={\bf q}_2 + {\bf q} $.  
Finally the annihilation contribution is
\begin{eqnarray}
\lefteqn{\hspace{-.10cm}
\frac{\langle \Psi \arrowvert H_{an} \arrowvert \Psi \rangle}
{\langle \Psi\arrowvert \Psi \rangle} =  \frac{9}{4}
{\cal{F}}_{an}^{\pm}\int d{\bf
q}_1 d{\bf q}_2 d{\bf q} F^*_{\mu_1 \mu_2 \mu_3}({\bf q}_1,{\bf q}_7)\hat{V}(q) }
\nonumber
\\
& &\hspace{-.4cm}
F_{\nu_1 \nu_2 \nu_3}({\bf q}_8,{\bf q}_2) 
\frac{(\omega_1+\omega_7)(\omega_2+\omega_8)}
{\sqrt{\omega_1\omega_2\omega_7\omega_8}}
D_{\mu_3 \nu_3}({\bf q}) D_{\mu_1 \alpha}({\bf q}_1)
\nonumber
\\
& & \hspace{2.5cm}
D_{\mu_2 \alpha}({\bf q}_7) D_{\nu_1 \beta}({\bf q}_8)
D_{\beta \nu_2}({\bf q}_2)  \ ,
\end{eqnarray}
with ${\bf q}_7 ={\bf q} - {\bf q}_1 $ and ${\bf q}_8 ={\bf q} - {\bf q}_2 $.
The first two contributions exactly cancel the infrared
singularity in the instantaneous potential.
The color factors ${\cal{F}}^{C=\pm}$
depend upon the  $C$ parity and follow
directly from the  expectation value of the density-density
($f^{abc}f^{aef}$) term in the Coulomb kernel: 
${\cal{F}}_{sc}^{+} = 
\frac{f^{abc}f^{aef}f^{bei}f^{cfi}}{f^{abc}f^{abc}} = \frac{-36}{24}
=-\frac{3}{2}$, 
${\cal{F}}_{sc}^{-} = 
\frac{d^{abc}d^{aef}f^{bei}f^{cfi}}{d^{abc}d^{abc}} = \frac{-20}{40/3} =
-\frac{3}{2}$, 
${\cal{F}}_{an}^{+} = 
\frac{f^{abc}f^{aef}f^{bci}f^{efi}}{f^{abc}f^{abc}} = \frac{72}{24} = 3$ and
${\cal{F}}_{an}^{-} = 
\frac{d^{abc}d^{aef}f^{bci}f^{efi}}{d^{abc}d^{abc}} = \frac{0}{40/3} = 0$,
using the orthogonality relation $d^{abc}f^{abc} = 0$ and  computer algebra
(FORM~\cite{Vermaseren:2000nd}).
The $H_{an}$ interaction  splits 
the $0^{-+}$ s wave glueball state from the $1^{--}$ and $3^{--}$ levels, all
calculated degenerate in  the work of Ref.~\cite{Hou:1982dy}. Also, annihilation is
not present in simple
$ggg$ potential models as well as all models for $gg$
states which is why it has not been studied previously in investigations of
the lightest $0^{-+}$ mass, involving p wave two-gluon states, calculated  between
2.1 GeV~\cite{Szczepaniak:1995cw} and 2.6 GeV~\cite{mp}.

The  variational calculation entails nine-dimensional integrals which
where performed using the Monte Carlo method with  the 
adaptive sampling algorithm VEGAS~\cite{Lepage:1980dq}.  In general, numerical
convergence was achieved with about  $10^5$ samples. 
A study
of the glueball mass sensitivity to both statistical and variational 
uncertainties yielded error bars at the 3 to 5\% level.

For comparison we also calculated oddball masses using a simpler nonrelativistic
constituent model Hamiltonian~\cite{Bicudo_frascati} 
\begin{equation}
 H_M = \sum_i\frac{{\bf q}_i^2}{2m_g} +  V_0+ \sum_{i<j}[
\sigma r_{ij} -{\alpha
\over r_{ij} } 
   + V_{ss}
{\bf S}_i \cdot {\bf S}_j ] , \hspace{-.25cm} 
\end{equation}
with $ r_{ij} = |{\bf r}_i - {\bf r}_j|$ and
parameters taken from the $q \bar{q}$ funnel potential:
$V_0 =$ -0.90 GeV, $\alpha =$ 0.27, $\sigma =$ 0.25  GeV$^2$.  The
gap equation
gluon mass value, $m_g=$ 0.8 GeV, is also used.
Exactly diagonalizing,
$H_M \Psi^{JPC} = M_{JPC} \Psi^{JPC}$, and only adjusting $V_{ss}= 0.085$ GeV to
optimize agreement with lattice, yields the predicted $J^{++}$ glueball
Regge trajectory, $\alpha^M_P = 0.23 t + 1.0$, consistent with the pomeron.
In the $gg$ glueball calculation an algebraic color
factor of 2 multiplies each bracketed term in the potential.
With these fixed parameters the $ggg$ oddball spectrum was then obtained by
exact diagonalization using Jacobi coordinates and an  expansion in a harmonic
oscillator basis.  Further details will be reported in a
future communication.
The glueball masses and quantum numbers investigated are listed in
Table~\ref{statetable}.  
Note that the
$H^g_{eff}$ ground state   for the
$ggg$ glueball is technically not an oddball but the $0^{-+}$ with mass  3900
MeV. Also, the model $ggg$ glueballs are 
heavier than $gg$ states (both having low excitations)
which is consistent with predictions deduced from the holographic dual theory of
QCD~\cite{deTeramond:2005su}.

\begin{table}[h]  
\caption{\label{statetable} Glueball  quantum numbers and  masses
in MeV.}
\begin{ruledtabular}
\begin{tabular}{|c|c|cccccc|}
\hline
 Model&$J^{PC}$& $0^{-+}$ & $1^{--}$ &$2^{--}$&  $3^{--}$ & $5^{--}$ & $7^{--}$ 
\\ \hline
this work &color    &$f$ & $d$ & $d$ &  $d$ & $d$ &$d$   \\
&$S$        &0 & 1  & 2& 3 & 3 & 3 \\
&$L$        &0 & 0  & 0 &0 & 2 & 4 \\ 
$H^g_{eff}$& & 3900 & 3950 &4150 & 4150 & 5050 & 5900 \\
$H_{M}$ && 3400& 3490 & 3660 & 3920 & 5150 & 6140 \\\hline
lattice~\cite{mp}&  &3640 & 3850  & 3930 &4130 & &  \\
lattice~\cite{Meyer:2004jc}& &3250 &3100 & 3550 &4150 & &  \\
Wilson loops~\cite{ks}&&3770 & 3490 & 3710 & 4030 & & \\
\hline
\end{tabular}
\end{ruledtabular}
\end{table}

\begin{figure}
\psfig{figure=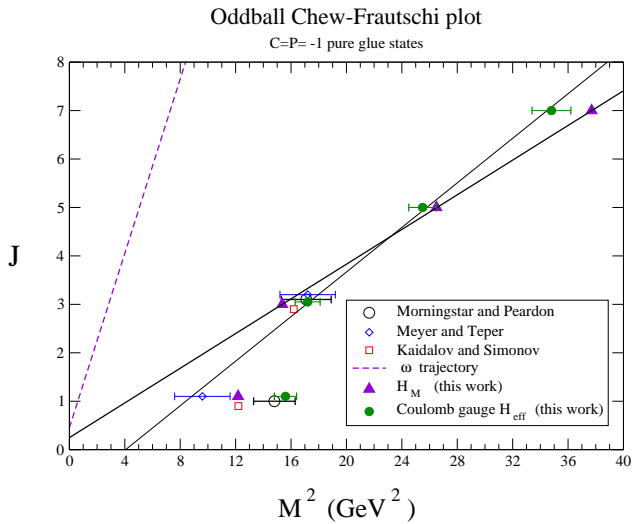,width=2.9in,angle=-90}
\caption{\label{odderonfig} Odderon trajectory 
from three-gluon spectroscopy and lattice compared to the  
$\omega$ meson Regge trajectory.}
\end{figure}

Figure \ref{odderonfig} summarizes our key findings 
and displays predicted oddball Regge trajectories from the alternative
approaches.  
Lattice results are represented by diamonds~\cite{Meyer:2004jc} and open
circles~\cite{mp}, while  boxes~\cite{ks} are constituent model predictions
using  a Wilson-loop inspired potential.  Solid circles and triangles
correspond
to the  $H^g_{eff}$ and $H_M$ models, respectively.
The resulting  odderon trajectories are represented by the solid
lines,  $\alpha_O^{eff} = 0.23 t - 0.88$  and 
$\alpha_O^{M} = 0.18t + 0.25$, whose differences provide an overall
theoretical uncertainty. The
dashed line is the
$\omega$ trajectory.

Several conclusions follow.  First,   this
work predicts  an odderon  having slope
similar
to the pomeron but   intercept even lower than
the $\omega$ value.  Second, the first odderon state 
is the $3^{--}$ and  not the $1^{--}$ which falls
on a daughter trajectory.  Unfortunately the other approaches did not report a
$5^{--}$  glueball which could confirm this point.  Since  at least two
points are necessary to establish a   trajectory we strongly recommend that
future studies calculate  higher $J$ states.
Lastly, there appears to be general model consensus that the
$3^{--}$ mass is close to 4 GeV and also support for the
$0^{-+}$ as a
ground state candidate for the $ggg$ system.

In summary, we have compared existing oddball predictions
to  our large-scale model evaluations for
$J^{--}$ glueballs. The results document an odderon
trajectory  subdominant to the pomeron  which can explain
its nonobservance in reactions with pomeron exchange.  Should
the odderon intercept be comparable to the $\omega$ value it may
be possible to see it in pseudoscalar~\cite{eb1} or tensor
meson~\cite{eb2} electromagnetic production  where the pomeron is absent.
However if the intercept is below 0.5, as we predict,
it is unlikely the odderon will be observed.

F.L. acknowledges a Fundacion del Amo-Univ. 
Complutense fellowship and the hospitality of the SLAC theory
group. Thanks to E. Abreu,  S. Brodsky, J. Vary and  P. Zerwas for useful
conversations. Research supported by Spanish Grant MCYT FPA 2004-02602 (F.L.)
and U. S. DOE Grant  DE-FG02-97ER41048 (S.C.).


\end{document}